Albert A. Michelson's Experimentum Crucis 1881 in Potsdam, Germany

Hans J. Haubold
Office for Outer Space Affairs, United Nations, Vienna, Austria, hans.haubold@gmail.com

Abstract. This paper reviews briefly the history of the Michelson experiment, invented and performed for the first time in the Astrophysical Observatory Potsdam in 1881. The paper draws attention to the International Michelson Colloquium, held from April 27 to April 30, 1981 in Potsdam (Germany). This paper is an attempt to reconsider a scientific event organized 40 years ago, as the follow-up to Einstein's Centenary in 1979, for Michelson's experiment done 140 years ago.

In his famous experiment in 1881 in the eastern basement of the Astrophysical Observatory Potsdam, Michelson intended to demonstrate that his interferometer was able to satisfy the task to verify the effect of the motion of the Earth on the propagation of light. It was expected that the velocity of light is composed of that of the Earth. The speed of the light should exceed the speed of the light which traverses the orbit by 30 km/s. In that case, the distance of the two images would depend on the orientation of the interferometer. The interferometer did not find any difference in the two velocities. Michelson had to conclude a so-called 'null result' that the propagation of light was determined by the walls, just as the propagation of sound in the air of the basement room had to relate to the walls. Michelson's interferometer result is a paradigmatic example of a null result in physics, a result may be said to be null when it not detected by the measuring device employed. The value returned by the measuring instrumentation is 'zero'. It is very rarely the case that an unadulterated zero result will occur since there will almost always be measurable, small interfering causes and resultant noise at play. Thus, a better description of a null result is that it is 'zero' plus small though annoying residual variations. Today, Michelson's original experiment and its many repetitions are considered as a venerable well understood historical chapter for which, at least from a physical point of view, there is nothing more to refine or clarify. Though, this is not necessarily true and this was also the subject of the Michelson Colloquium and it remains the subject until today [3, 4].

From April 27 to April 30, 1981, an international colloquium in honour of the physicist and first American Nobel laureate Albert Abraham Michelson (1852-1931) and his scientific work took place in Potsdam, Germany. The occasion for this Michelson Colloquium at the Astrophysical Observatory Potsdam was the centenary of the year in which the famous Michelson experiment was performed for the first time in the Astrophysical Observatory Potsdam [1].

The Michelson Colloquium was held under the auspices of the Academy of Sciences of the GDR (Central Institute for Astrophysics); its arrangement at the Astrophysical Observatory Potsdam was sponsored by the Einstein Laboratory for Theoretical Physics, Caputh, the Physical Society of the GDR, Berlin, and the Department of Physics of the Humboldt University, Berlin.

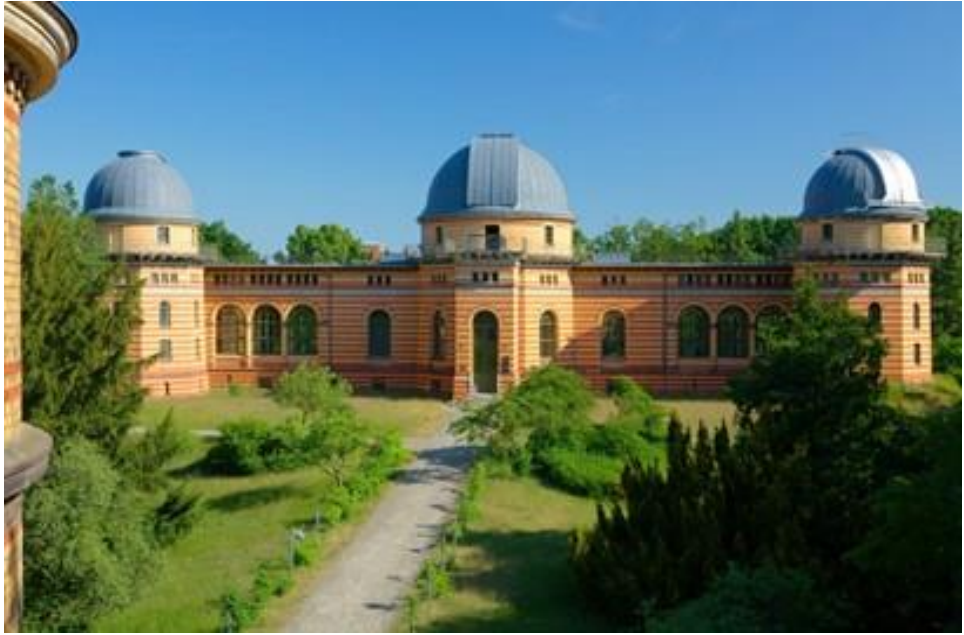

Fig. 1. Main building of the Astrophysical Observatory Potsdam.

About one hundred years ago, in April 1881, Michelson performed for the first time his interferometer experiment to determine the velocity of the Earth relative to the hypothetical luminiferous aether- an experiment which entered into history of physics and astronomy. The null-result of the experiment, rejecting the aether hypothesis of Fresnel, turned out to be fundamental for the evolution of physics, as a landmark on the way of the genesis of Einstein's theory of relativity. The incitement for undertaking such a crucial experiment Michelson had found in ideas of J.C. Maxwell. The interferometer by which he carried out his experiment and which he later on used in various investigations in physics and astronomy, Michelson invented during a visit to Europe, beginning at the end of 1880. The device was made by the optical firm Schmidt and Haensch in Berlin and the experiment was prepared at the Physical Institute of the Friedrich Wilhelms University of Berlin (later Humboldt University), located at the Reichstagsufer and lead by Hermann von Helmholtz. However, because of the sensitivity of the instrument against vibrations, one had to look for a place of lower vibration level. The memorable experiment was finally realised in the Astrophysical Observatory Potsdam, not far from Berlin. As Michelson later mentioned, the then director of the Observatory, H.C. Vogel, was at once interested in the experiment. The whole experimental set up was placed in the basement of the east dome of the main building of the Observatory (Fig. 1).

The lectures delivered at the Michelson Colloquium, with large thematic variety gathering round the Michelson experiment as the focus, especially appreciated its importance for physics and astronomy and dealt with philosophical and scientific historical aspects of the Michelson experiment. They took place during two days in a solemnly decorated room of the old city-hall of Babelsberg.

On April 28, H.-J. Treder (1928-2006) opened the jubilee colloquium and Michelson exhibition by welcoming special guests and participants to the internationally organized colloquium which was understood as a natural follow-up to the Einstein Centenary celebrations held in Berlin and Potsdam in 1979 [5].

Subsequently Treder asked J. Auth (1930-2011; Humboldt University, Berlin, Germany) for delivering the opening lecture "Albert A. Michelson at the University of Berlin" [1]. By means of documents from the archive of the Humboldt University, Auth portrayed Michelson's scientific visit to Berlin and the preparation of the experiment. In that he built upon a detailed investigation of H.J. Haubold and R.W. John "Albert A. Michelson's aether drift experiment 1880/1881 in Berlin and Potsdam" [2, 6]. Auth also sketched the social life of that time at the University of Berlin. In great detail he devoted his attention to the theory and the physical consequences of the Michelson experiment.

This general view over Michelson's visit to Berlin was followed by the celebration lecture of Dorothy Michelson Livingston (1906-1994; New York, USA) "Michelson and Einstein, artists in science" [1]. Mrs. Michelson Livingston (Figs. 2, 3 and 8) drew a vivid picture of her famous father and analysed the characteristics of his creative activity, especially in comparison with Albert Einstein. From her memories she made the audience familiar with the great experimental physicist Albert A. Michelson, which was painting with pleasure in his spare time and was athletically active up to the old age. At that the lecturer showed slides from her private photo collection. Mrs. Michelson Livingston is the author of the outstanding Michelson biography "The Master of Light", first published in 1973 [7]. She finished her lecture with words spoken by Einstein in appreciating Michelson on the occasion of meeting him at the California Institute of Technology, Pasadena, in January 1931, a few months before the death of Michelson: "It was you who led the physicists into new paths, and through your marvelous experimental work paved the way for the development of the theory of relativity".

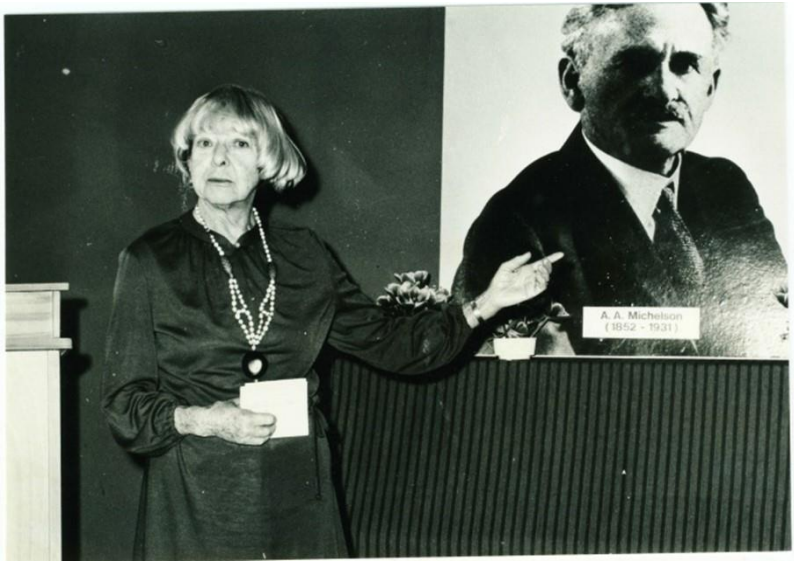

Fig. 2. Dorothy Michelson Livingston opening speech of the Michelson Colloquium.

The next lecturer announced was R.S. Shankland (1908-1982; Case Western Reserve University, Cleveland, Ohio, USA) "Michelson in Potsdam" [1]. It happened in Cleveland in 1887, when Michelson in cooperation with E.W. Morley repeated the aether-drift experiment with higher precision. The result confirmed the null-result of the Potsdam experiment. Shankland himself was leading engaged in the disclosure of the misinterpretation of the result announced by D.C. Miller in single carrying out a further repetition of the Michelson experiment. Shankland lucidly and concisely presented that part of Michelson's scientific activity, which begins with his collaboration with S. Newcomb, then, in 1880, lead to the visit to Europe – stations were Paris and Berlin/Potsdam – and which included the invention of the Michelson interferometer, its construction in Berlin, the preparation of the experiment in von Helmholtz' institute, and finally, the realization of the experiment in the Astrophysical Observatory Potsdam. At that time, H. von Helmholtz was already a famous scientist, inter alia for his contributions to physical optics and the foundation of physiological optics. His laboratory was a renown modern centre for optical research. This intensive scientific environment no doubt was an important factor in Michelson's progress, as Shankland stated.

After this lecture a short annotation given by H. Melcher (b.1927; Training College for Teachers, Erfurt, Germany) followed, concerning some special questions of the history of the aether-drift experiment and the genesis of special relativity [1, 8].

The afternoon lectures began with a joint contribution by R. Rompe (1905-1993; Physical Society, Berlin, Germany) and G. Albrecht (1930-2015; German Academy of Sciences, Berlin, Germany) "The importance of experiments for the progress in physics" [1]. Rompe pointed out, how "the nearness of an experiment to experience", what was just existing yet in Michelson's investigations, today threatens to fade away due to the more and more increasing complexity of the experimental proceeding in physics and the necessary inclusion of electronic data processing. He stressed the tight nexus of physical conception, mathematical theory, and experiment. So, on one hand, the experimental advance is essential for the development of physical theory, but, on the other hand, according to a remark made by Einstein to Heisenberg, designing an experiment is again decisively co-determined by the theory. The lecturers spoke about the methodical benefit one may also yet today derive from Albert A. Michelson's, the great master of precision optics, style of working. A further part of the lecture, separately read by K. Lanius (1927-2010; Institute of High Energy Physics, Zeuthen, Germany) was dealing with the statistical evaluation of experiments [1].

After that H.-G. Schoepf (1928-2004; University for Science and Technology, Dresden, Germany) delivered his lecture "Maxwell's aether theories" [1]. In an enthusiastic and exciting manner, he coherently displayed the ways which have led Maxwell via different mechanical aether models to his famous equations of electromagnetism.

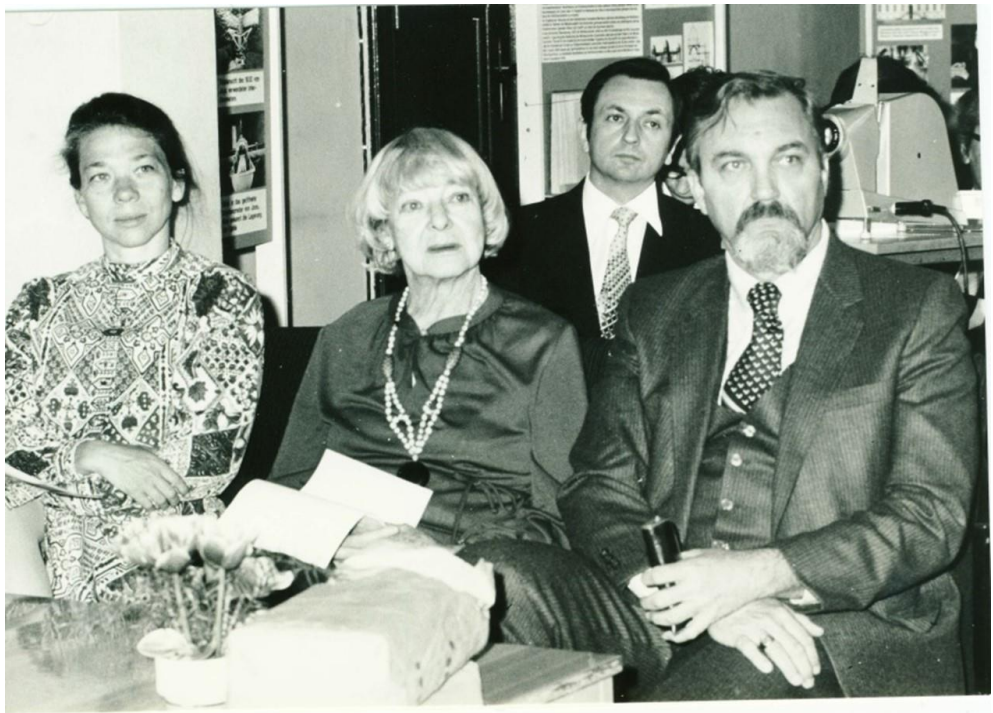

Fig. 3. From left U. Ulmer (Youngest daughter of D. Michelson Livingston), D. Michelson Livingston, R.W. John (1942-2007; co-organizer of the Colloquium), L.S. Swenson.

The scientific programme of the first day was concluded by the lecture of the well-known historian of science, L.S. Swenson Jr. (1932-2016; University of Houston, Texas, USA): "The Michelson-Morley-Miller experiments and the Einsteinian synthesis" [1, 9]. Swenson (Fig. 3) described the history of these experiments in connection with the rise of the special theory of relativity. In that he stressed that the history of physics must be seen in connection with the history of technology. At the end of his lecture Swenson raised several questions concerning the history of the theory of relativity. He also highlighted that in the Munich Deutsches Museum letters from the Einstein – von Laue correspondence are deposited which should be published in the future.

The session of the second day of the Michelson Colloquium started with Vigier's (1920-2004; Institute Henri Poincare, Paris, France) lecture, "Einstein's and Dirac's aether models" [1, 10]. After having stated the classical aether models are killed by the null-result of the Michelson experiment,

Vigier (Fig. 4) investigated the justification of stochastic covariant aether models. He dealt with the relativistic generalization of results obtained by Einstein and von Smoluchovsky on Brownian motion. Discussing the relation of superluminal velocities and causality he rested upon experiments performed by A. Aspect.

In the following lecture, "Interference and correlation phenomena in quantum theory", Kaschluhn (1927-1994; Institute of Theoretical Physics, Humboldt University, Berlin, Germany) perspicuously worked out a problem connected with the microphysical description of single systems, the overcoming of which obviously demands additional assumptions about the quantum mechanical measuring process [1].

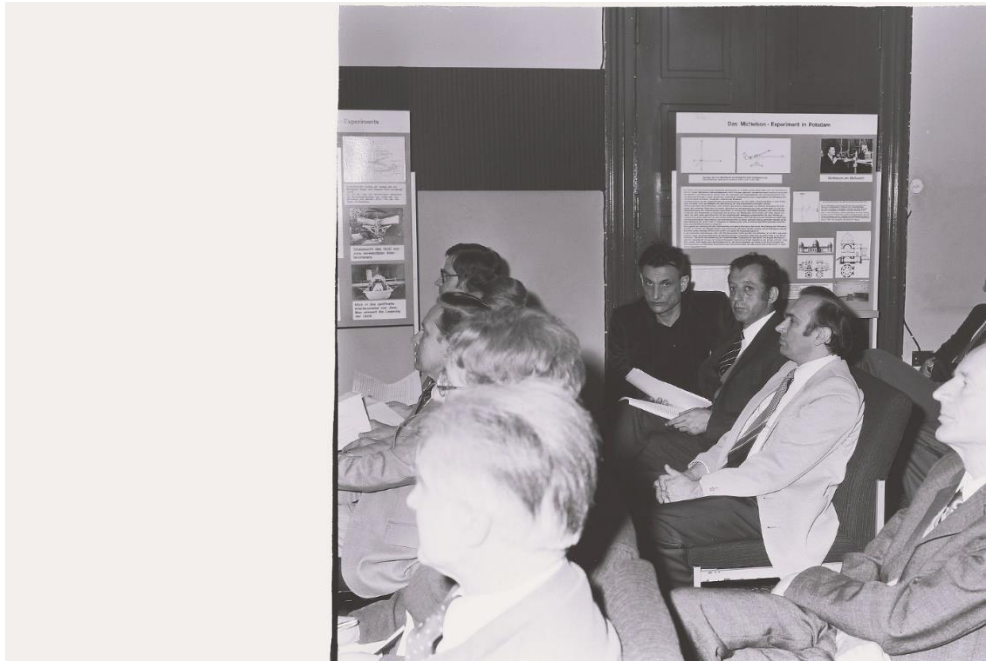

Fig. 4. V. Vigier, Z. Maric, and E. Kreisel (b. 1937). In the background an example is visible of the 12 large exhibition panels produced for the Einstein centenary in Potsdam and Berlin and the Michelson Colloquium in Potsdam.

Z. Maric (1931-2006; Institute of Physics, Belgrade, Yugoslavia) posed his lecture, likewise dealing with quantum field theory, under the theme: "Vacua and symmetries". Maric (Fig. 4) investigated the vacuum concept on different stages of development of physics, especially he emphasized the relation of this concept to the internal geometry of space.

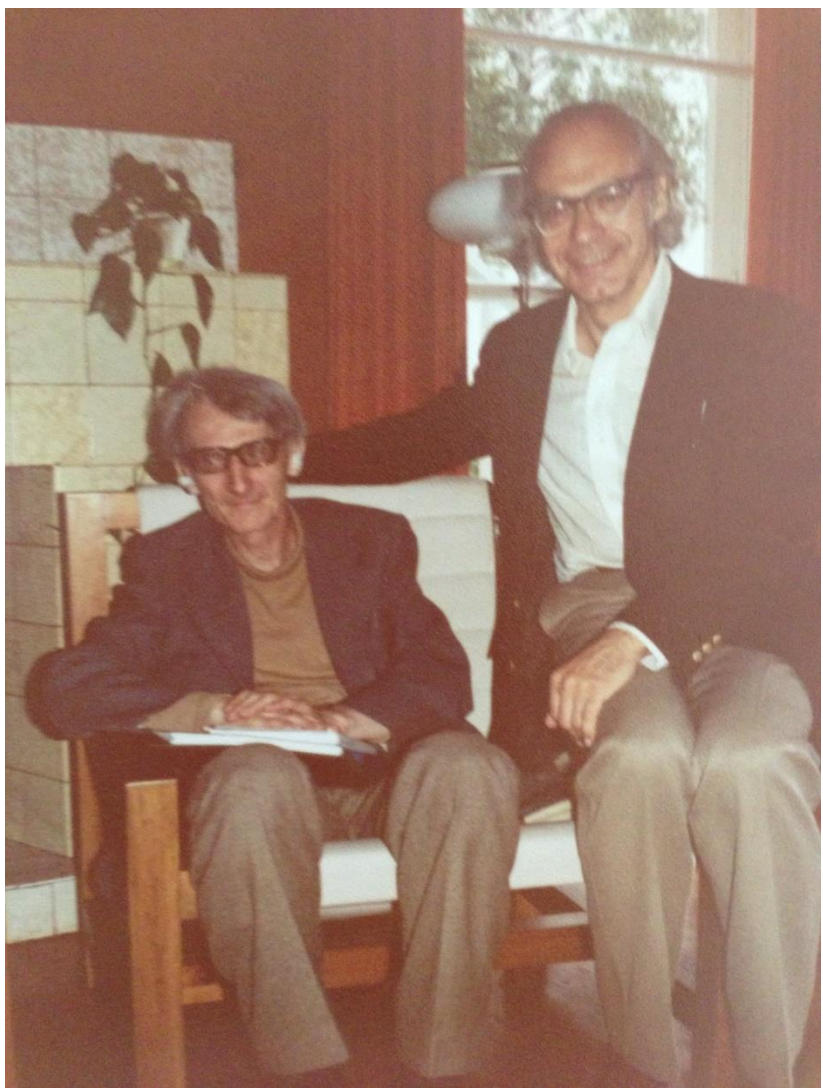

Fig. 5. H.-J. Treder and J. Stachel in front of the fireplace in Einstein's Summer House in Caputh.

    J. Stachel (b.1928) focused in his lecture titled "Einstein and Michelson: The Context of discovery and the context of justification" on the fact that a philosopher of science is not interested in the context of discovery, but in the context of justification and that it appears amazing to what extent the logical analysis of relativity coincides with the original interpretation by its author, as far as it can be constructed from the scanty remarks in Einstein's publications [1, 11]. According to Stachel (Fig. 5), in contradistinction to some developments in quantum theory, the logical schema of the theory of relativity corresponds surprisingly with the program which controlled its discovery. This agrees with the independent analysis of philosophers and scientists as Hans Reichenbach and Gerald Holton [12]

    The final lecture of the celebration colloquium was given by H.-J. Treder (1928-2006; Astrophysical Observatory Babelsberg, Potsdam, Germany): "The Michelson experiment as experimentum crucis" [1]. Dealing with this theme, and simultaneously aspiring a synopsis of the most important statements involved in the foregoing lectures, Treder (Fig. 5) stressed the following central aspects: The art and importance to perform experiments, the creation of new experiments from putting of important physical questions; the question of the existence or non-existence of the aether; the meaning of classical questionings, from Michelson to Einstein and from Einstein to Bohr, at the present time.

    The organizers of the Michelson Colloquium took advantage of this event in planning visiting opportunities between lecture sessions and beyond. All participants of the colloquium were invited to

visit the main building of the Astrophysical Observatory Potsdam (Fig. 1), the Einstein Tower on the Telegrafenberg (Fig. 6), and the Einstein House in Caputh near Potsdam (Fig. 7). Many of the colloquium lectures did touch historical, scientific, and social issues which are part of research programmes as analysed in depth, for example, in Holton's scientific papers [12]. Long time after the discoveries addressed by lectures of the Michelson Colloquium the decision was made to rename the main building of the Astrophysical Observatory to be the Michelson building to honour Michelson's first experiment.

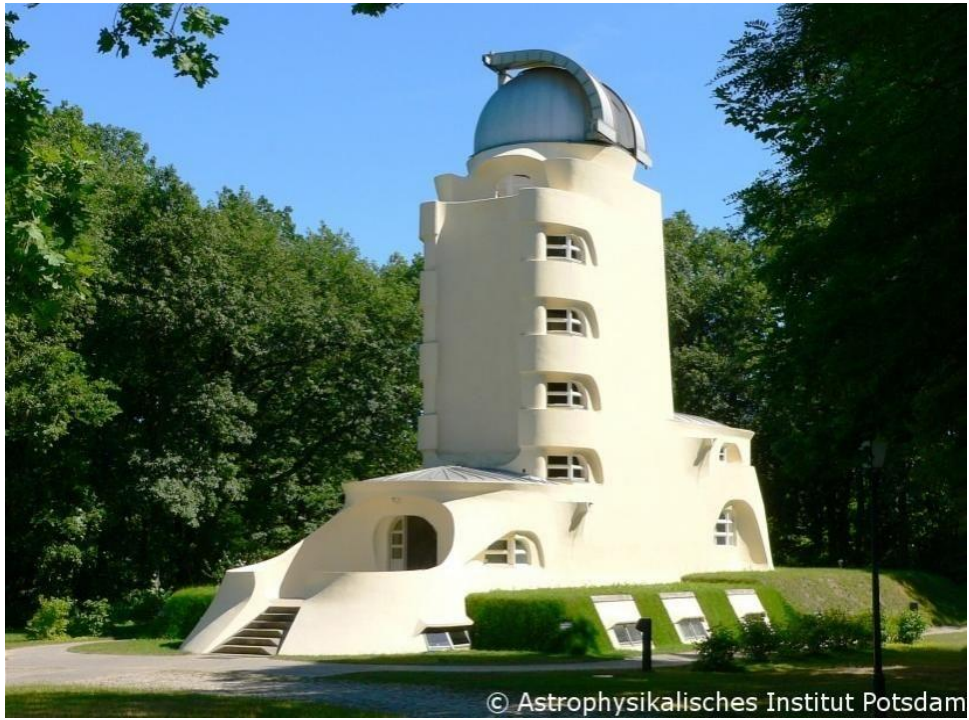

Fig. 6. Einstein Tower, Telegrafenberg, Potsdam.

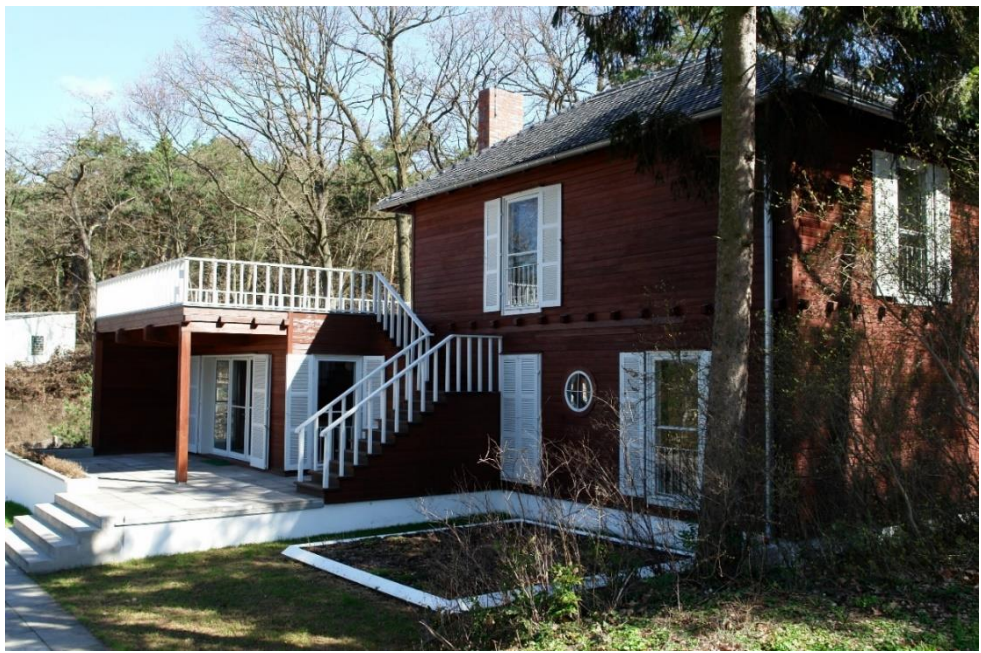

Fig. 7. Einstein's Summer House in Caputh.

A number of participants had the unique opportunity to be accommodated as guests in the summer house of Einstein in Caputh and were invited to continue the colloquium deliberations in guided tours and discussion sessions (Fig. 8). The summer house built for Albert Einstein in Caputh close to Potsdam in 1929 by the young architect Konrad Wachsmann is of great cultural and architectural importance. The house is the result of close interaction between the requests of Albert Einstein and the ideas of Konrad Wachsmann. The ample window front of the spacy living room with its open lightness goes back to the ideas of the architect and invited even at Einstein's time to meeting sessions. Unfortunately, constant change of use and ownership as well as lack of financial resources and material resulted in a lack of maintenance work. Over long time, legal debate about the ownership of the house prevented necessary restoration. Yet through all ups and downs of its history the house has preserved a high degree of authenticity and has always remained the »summer house of Albert Einstein«. During the Michelson Colloquium the house was elevated to be the host of the Einstein Laboratory for Theoretical Physics lead by H.-J. Treder.

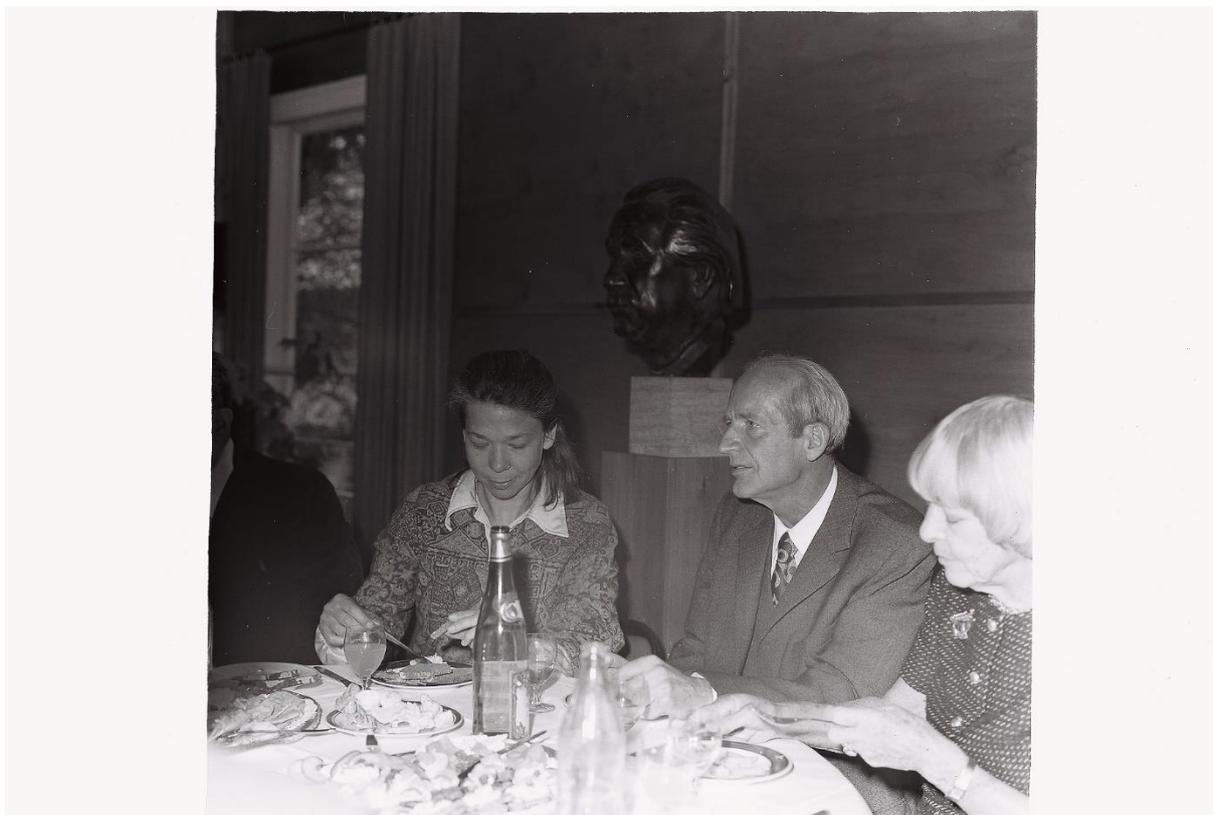

Fig. 8. U. Ulmer, F.W. Jaeger (1914-2000; former director of the Einstein Tower, Telegrafenberg, Potsdam), D. Michelson Livingston in the Einstein House Caputh, in the background the Einstein bust created by Heinrich Drake in 1981.

A unique element of the programme of the Michelson Colloquium for all participants was the opportunity to visit the Einstein Tower through guided tours and discussion sessions. The Einstein Tower is an astrophysical observatory on the Telegrafenberg later to become the Albert Einstein Science Park in Potsdam, Germany, built by architect Erich Mendelsohn. It was built on the summit of the Potsdam Telegrafenberg to house a solar telescope designed by the astronomer Erwin Finlay-Freundlich. The telescope supports experiments and observations to validate (or disprove) Albert Einstein's relativity theory. The building was first conceived around 1917, built from 1919 to 1921, and became operational in 1924. Although Einstein never worked there, he supported the construction and operation of the telescope. Light from the telescope is directed down through the shaft to the basement where the instruments and laboratory are located. In 1911 Einstein published the initial version of his General Theory of Relativity. One of the predicted effects according to the theory was a slight shift of spectral lines in the sun's gravitation field, now known as the red shift. The solar observatory in Potsdam was designed and constructed primarily to verify this phenomenon.

100 years were gone since the Michelson experiment was performed for the first time in Potsdam – this also meant 50 years since Albert A. Michelson's death – but, as the lectures of the Michelson Colloquium showed, the physical and scientific historical discussion continuously inflames at this decisive experiment which signalized a revolution in the development of physics. Also, in dealing with questions in more distant fields, the discussions looked back at this experiment as a pioneering event in the history of physics.

Acknowledgements

The author gratefully acknowledges the support and advice provided for this paper over a long period of time by Reiner Wilhelm John, Hans-Juergen Treder, Johann Wempe, Dorothy Michelson Livingston, Loyd S. Swenson, Robert S. Shankland, Gerald Holton, and William Fickinger.

References

[1] H.-J. Treder (Editor), Proceedings of the Michelson Colloquium, Potsdam, 28-29 April 1981, Potsdam, Germany, Astronomische Nachrichten 303 (1982) 1-96.
[2] S. Goldberg and R.H. Stuewer (Editors), The Michelson Era in American Science 1870-1930, AIP Conference Proceedings 179, American Institute of Physics, New York 1988. W. Fickinger and K.L. Kowalski (Editors), Modern Physics in America: A Michelson-Morley Centennial Symposium, AIP Conference Proceedings 169, American Institute of Physics, New York 1988.
[3] M. Consoli and A. Pluchino, Michelson-Morley Experiments: An Enigma for Physics and the History of Science, World Scientific, Singapore 2019.
[4] A. Franklin and R. Laymon, Measuring Nothing, Repeatedly Null Experiments in Physics, Morgan & Claypool, 2019.
[5] H.-J. Treder (Editor), Einstein-Centenarium 1979, Akademie-Verlag, Berlin 1979. Albert Einstein Akademie-Vortraege, Wiederabdruck durch die Akademie der Wissenschaften der DDR, Akademie-Verlag, Berlin 1979.
[6] H.J. Haubold and R.W. John, NTM-Schriftenreihe fuer Geschichte der Naturwissenschaften, Technik und Medizin (Leipzig) 19 (1982) 31-45.
[7] D. Michelson Livingston, The Master of Light – A Biography of Albert A. Michelson, Charles Scribner's Sons, New York 1973; published in Kindle format by Plunkett Lake Press in 2021. http://hdl.handle.net/2186/ksl:2006061209
[8] J. Renn, L. Divarci, P. Schröter, A. Ashtekar, R.S. Cohen, D. Howard, S.S. Abner Shimony (Editors), Revisiting the Foundations of Relativistic Physics: Festschrift in Hounor of John Stachel, Springer Netherlands 2003.
[9] L.S. Swenson, Jr., The Ethereal Aether: A History of the Michelson-Morley-Miller Aether-Drift Experiments 1880-1930, University of Texas Press, Austin & London 1972. https://iopscience.iop.org/book/978-1-64327-738-7/chapter/bk978-1-64327-738-7ch7
[10] V. Vigier, Relativistic Interpretation (with Non-Zero Photon Mass) of the Small Ether Drift Velocity Detected by Michelson, Morley and Miller Apeiron, 1997, Volume 4, No. 2-2, pp. 71-76.
[11] J. Stachel, Einstein from 'B' to 'Z', Birkhäuser, Boston 2002.
[12] G. Holton, Thematic Origins of Scientific Thought: Kepler to Einstein, revised edition, Harvard University Press, Cambride 1988. German edition by Suhrkap, Frankfurt am Main 181 titled Thematische Analyse der Wissenschaft: Die Physik Einsteins und seiner Zeit.